\documentclass[twocolumn,showpacs,amsmath,amssymb,prb,floatfix]{revtex4}

\usepackage{graphicx}
\usepackage{bm}

\newcommand{\pderiv}[2]{\frac{\partial #1}{\partial #2}}
\newcommand{\half}{\frac{1}{2}}
\newcommand{\units}[1]{\ \mathrm{#1}}
\newcommand{\BohrMag}{\mu_\mathrm{B}}

\begin{document}

\title{Single-qubit gates and measurements in the surface
acoustic wave quantum computer}
\date{\today}
\author{S.~Furuta}
\altaffiliation[Also affiliated to: ]{Centre for Quantum
Computation, Department of Applied Mathematics and Theoretical
Physics, Wilberforce Road, Cambridge CB3 0WA, UK} \email[email:
]{suguru.furuta@qubit.org}
\author{C.~H.~W.~Barnes}
\affiliation{Cavendish Laboratory, Department of Physics,
University of Cambridge, Madingley Road, Cambridge, CB3 0HE, UK.}
\author{C.~J.~L.~Doran}
\affiliation{Cavendish Laboratory, Department of Physics,
University of Cambridge, Madingley Road, Cambridge, CB3 0HE, UK.}

\begin{abstract}
In the surface acoustic wave quantum computer, the spin state of
an electron trapped in a moving quantum dot comprises the physical
qubit of the scheme. Via detailed analytic and numerical modeling
of the qubit dynamics, we discuss the effect of excitations into
higher-energy orbital states of the quantum dot that occur when
the qubits pass through magnetic fields. We describe how
single-qubit quantum operations, such as single-qubit rotations
and single-qubit measurements, can be performed using only
localized static magnetic fields. The models provide useful
parameter regimes to be explored experimentally when the
requirements on semiconductor gate fabrication and the
nanomagnetics technology are met in the future.
\end{abstract}


 \maketitle

\section{Introduction}
\begin{figure}
 \centering
 \includegraphics[width=8.6cm]{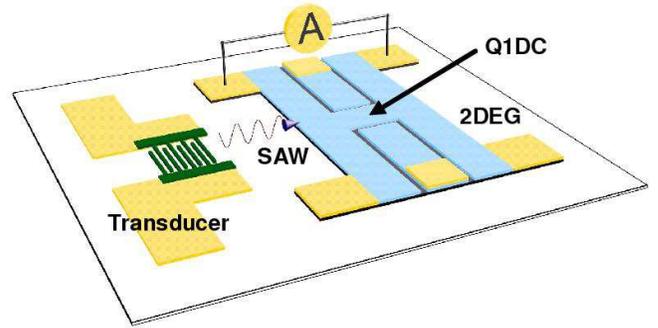}
 \caption{
 (Color online) Schematic diagram of an experimental
  device for producing quantized
 acoustoelectric currents through a narrow Q1DC constriction.
 \label{sawdevice}
 }
\end{figure}

Quantum computation promises enormous technological advances in
the field of information
processing\cite{{Feynman:1982},{Shor:1994},{Ekert:1996},
{NielsonChuang:Book:2000}} and the quest for its realization has
attracted many strong contenders in the field of physics and
engineering. This paper is concerned with a scheme for quantum
computation put forward by Barnes, Shilton and
Robinson,\cite{Barnes:PRB:2000} which falls into the semiconductor
quantum dot
category.\cite{Loss:PRA:1998,Burkard:PRB:99,Loss:Proc:1999,
Troiani:PRB:2000,DiVincenzo:Nat:2000,Elzerman:PRB:2003,Calarco:PRA:2003}
The proposal for quantum computation is based on the results of
ongoing experiments that have demonstrated the capture and
transport of single electrons in moving quantum
dots.\cite{Shilton:96,{Talyanskii:97},Talyanskii:98} The dots are
formed when a surface acoustic wave (SAW) travels along the
surface of a piezoelectric semiconductor containing a
two-dimensional electron gas (2DEG). See Fig.~\ref{sawdevice} for
a schematic diagram of the device. When the SAW is made to pass
through a constriction in the form of a quasi-one-dimensional
channel (Q1DC), the induced piezoelectric potential drags
electrons into and along the Q1DC. In certain parameter regimes
the device transports one electron per potential minimum of the
SAW.\cite{Talyanskii:97} The spin on the trapped electron
represents the physical qubit. Quantum computation involves
performing qubit operations on the trapped electrons as they move
with the speed of the SAW.

Many schemes for quantum computation, such as conventional quantum
dots,\cite{Loss:PRA:1998,{Vrijen:2000}} doped
silicon,\cite{Kane:1998} superconducting boxes,\cite{Makhlin:2001}
and ion traps,\cite{Wineland:2003} involve static qubits. The
surface acoustic wave quantum computer on the other hand is of the
``flying qubit'' type, which include linear optics
schemes,\cite{knill:2001} some ion trap schemes with ion
shuttling\cite{Cirac:Nature:2000} and schemes based on coherent
electron transport in quantum
wires.\cite{Bertoni:PRL:1999,Barnes:PRB:2000,APopescu:2003}  All
these have in common that the carriers of quantum information
physically move through space during the computation. Flying
qubits have the advantage of being able to distribute information
quickly over large distances across the quantum circuit when
decoherence times are short and to interface with quantum memory
registers at fixed locations. Another advantage of the SAW quantum
computation scheme is its ensemble nature. It intrinsically
performs time-ensemble computation, in much the same way NMR
quantum computation performs molecular-ensemble
computations.\cite{Cory:Proc:1997,chuang:1998,Jones:Proc:2003}
Time-ensemble computation alleviates the demand for single-shot
spin measurements and has the advantage of being robust against
small random errors.

This paper considers proposals\cite{Barnes:PRB:2000} for
implementing quantum gates on single SAW qubits using only static
magnetic fields generated by surface magnetic gates. Detailed
modeling of the gate operation has been accomplished by means of
both analytic solutions and numerical simulations of the Pauli
equation. We show how electrostatically confined moving electrons
behave under the influence of various magnetic fields and discuss
the implications for quantum computing with surface acoustic wave
electrons.

The physics of the SAW-guided qubit is explained in more detail in
section \ref{Section:Qubit}. In sections \ref{Section:Unitary} and
\ref{Section:Readout}, we present results on single-qubit unitary
gates and single-qubit readout gates based on the Stern-Gerlach
effect. Section \ref{Section:Neglect} discusses some of the
decoherence processes involved in quantum-dot based schemes.
Section \ref{Section:Summary} is a summary of the results with
parameter regimes of interest for future experiments.

\section{Scheme for quantum computation}\label{Section:Qubit}
We begin by summarizing the quantum computation scheme proposed by
Barnes, Shilton and Robinson.\cite{Barnes:PRB:2000} Figure
\ref{sawdevice} shows a schematic diagram of the experimental
setup originally designed to demonstrate quantized currents in
semiconductors.\cite{Shilton:96,{Talyanskii:97},Talyanskii:98} A
NiCr/Al interdigitated transducer is patterned on a GaAs/AlGaAs
heterostructure. A narrow depleted Q1DC splits the 2DEG into two
regions, the source and drain. When a high frequency AC signal is
applied to the transducer, a SAW propagates through the 2DEG,
producing a periodic piezoelectric potential across the 2DEG. The
potential drags electrons in the source region through the narrow
Q1DC constriction into the drain. It has been shown experimentally
that over a range of SAW power and gate voltages, the current
passing through the Q1DC is quantized in units of $ef$, where $e$
is the electronic charge and $f$ is the frequency of the
SAW.\cite{Shilton:96,Talyanskii:97,Talyanskii:98} The smallest
quantized current observed corresponds to the transport of a
single electron in each SAW minimum. Typically, the SAW in GaAs
moves at $2700\units{ms}^{-1}$ at a frequency of around
$2.7\units{GHz}$, with an applied power of
$3-7\units{dBm}$.\cite{Talyanskii:97} These parameters produce
currents in the range of nanoamps.

\begin{figure}
 \centering
 \includegraphics[width=8.6cm]{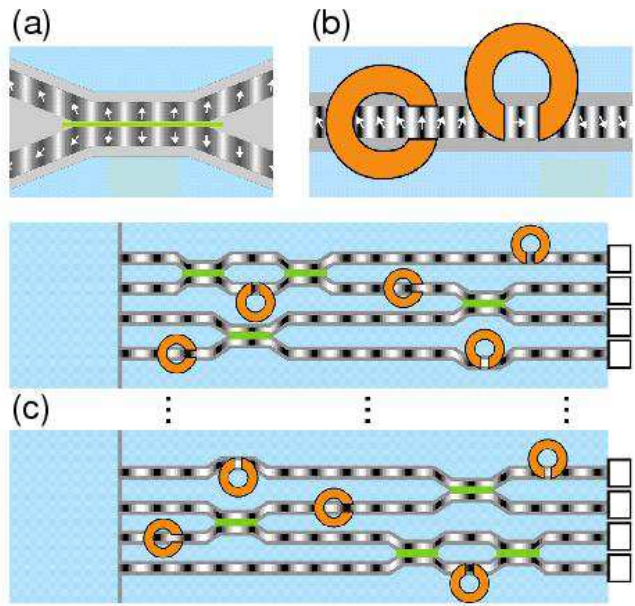}
 \caption{
 (Color online) Schematic diagram of a quantum gate network in a SAW quantum
 computer: (a) Two-qubit tunneling gates;
 (b) One-qubit magnetic gates (in various orientations);
 (c) Gate network for quantum computation with SAW electron spins.
  Gray lines running horizontally represent Q1DCs; blackened
  regions
 indicate the SAW minima where the qubits reside;
  arrows represent spin polarization;
  rings represent magnetic surface
 gates; white squares represent readout gates.
 \label{sawnetwork}
 }
\end{figure}

Given the ability to trap single electrons in the SAW minima, the
scheme for quantum computation is as follows. It is possible for
an array of $N$ Q1DCs in parallel to capture $N$ qubits in every
$M$th minimum, with a single electron in each Q1DC, producing a
qubit register along the SAW wavefront. $M$ can be chosen
sufficiently large to ensure that the Coulomb interaction between
successive qubit registers do not interfere with each separate
computation. The qubits move with the minima of the SAW, passing
through a sequence of static one- and two-qubit gates before
arriving at an array of spin readout devices. One-qubit gates may
be operated by nanoscale electromagnetic fields. Where two-qubit
gates are needed, neighboring Q1DCs are allowed into a tunnel
contact controlled by a potential on a surface
gate.\cite{Barnes:PRB:2000} The use of the Coulomb coupling
between neighboring Q1DCs is a common tool in spintronics which
can be used to generate entangled states in dual-rail qubit
representations.\cite{Ionicioiu:PRA:2001,Kitagawa:PRL:1991}
Figure~\ref{sawnetwork} illustrates the network of Q1DCs and qubit
gates envisaged for performing a particular quantum computation.

The SAW-trapped electron is well confined in all three spatial
directions. In this paper we define Cartesian coordinates such
that the SAW propagates along the $x$ axis with the $z$ axis
normal to the 2DEG. The 2DEG is produced by a band-energy mismatch
at the GaAs/AlGaAs interface which gives rise to a confining
potential in the $z$ direction. The energy level spacings in the
2DEG well are on the order of
$50-100\units{meV}$.\cite{Kelly:Book:95} Further confinement in
the $y$ direction is provided by an extended Q1DC, etched into the
surface so as to avoid screening the SAW-induced potential with
metallic surface gates.\cite{Utko:2003} Finally, the confinement
in the $x$ direction is due to the SAW potential minimum which is
approximately sinusoidal. The SAW amplitude is typically 40
meV,\cite{Robinson:PRB:2001} which is sufficiently large to
prevent qubits being lost via tunneling into neighboring SAW
minima.

There are two important aspects of the SAW quantum computation
scheme that distinguishes it from other similar quantum dot
schemes. First, the scheme provides repetitions of the same
quantum computation with each passing of a single wavefront of the
SAW. Therefore, a statistical time-ensemble of identical
computations can be read out at the end of the Q1DCs as a
measurable current, alleviating the need for single-electron
measurements. Two sources of noise in the measured current may be
estimated as follows: The shot noise is largely determined by how
well the current is quantized to $I=e f_\mathrm{SAW}$ and
precisions of $< 0.1\%$ can be experimentally
achieved.\cite{Robinson:Proc:2003} Johnson noise arises from the
resistance of the ohmic contacts and the 2DEG which are on the
order of $\sim 100-1000\units{\Omega}$. At temperatures of
$1\units{K}$ they produce a rms voltage noise spectral density on
the order of $10^{-10}\units{V/\sqrt{Hz}}$ at most. This cannot
drive a current noise through the SAW device since its effective
intrinsic impedance of $10\units{M\Omega}$ is comparatively very
large. See experimental papers\cite{Shilton:96,Talyanskii:97,
Talyanskii:98,Robinson:PRB:2002,Robinson:Proc:2003} for more
detailed discussions. The second key aspect of the scheme is the
static nature of the gate components of the quantum circuit. This
alleviates the need for strong, targeted and carefully-timed
electromagnetic pulses that can be difficult and expensive to
implement. The requirement of such expensive control resources
often limit the scalability of most quantum computing
implementations.

It would certainly be convenient, though not essential, to have a
means of preparing a pure fiducial qubit state. In NMR quantum
computing, operations are carried out on ensembles of replica
qubits which remain close to a highly mixed state of thermal
equilibrium. Nevertheless, a successful readout of the computation
is obtained because the sum over many identical computations
provides a measurable signal. Similarly in the SAW quantum
computing scheme, states close to the maximally-mixed state are
still useful because of the time-ensemble nature of the scheme.
However, in contrast to NMR schemes, it is in principle possible
to read out single electrons in the SAW scheme.  We will therefore
begin with nonensemble quantum computation in mind and only later
exploit the advantages of ensemble computation to deal with noise.
Of course, the more pure the qubit states remain, the faster the
computation will converge to the result. For these reasons, we
describe in this paper two simple methods for preparing pure
fiducial qubit states.

\begin{figure}[!t]
 \centering
 \includegraphics[width=8.6cm]{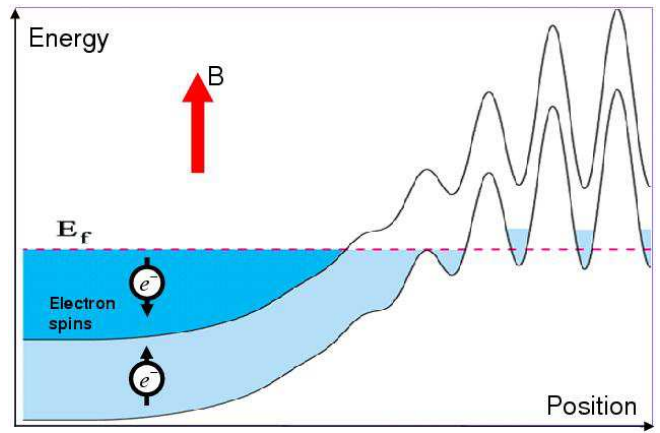}
 \caption{
 (Color online) Energy diagram showing
 polarized electron capture by means of Zeeman band splitting
 in the presence of a uniform magnetic field:
 Arrows on electrons indicate spin polarizations;
 $E_f$ is the Fermi energy. The SAW is strongly screened in the
 bulk
 2DEG below the Fermi energy. As the SAW enters the Q1DC
 constriction, the confining potential begins to form.
 The probability of capturing spin-down
 is small at the point the minority spin-type dot
 forms (upper curve). In contrast, the probability of capturing
 spin-up electrons is high when the majority spin-type dot forms
 (lower curve).
 \label{BCapture}
 }
\end{figure}

In the original proposal for the SAW quantum
computer,\cite{Barnes:PRB:2000} it was noted that the application
of an external magnetic field of about $1\units{T}$ will influence
which spin states of electrons are favored in the capture
process.\cite{Robinson:PRB:2001} This polarized capture process
can be summarized as follows. The SAW is strongly screened in the
bulk 2DEG until it is raised above the Fermi energy and the
quantum dot begins to form. When the higher-energy minority
quantum dot forms for the higher-energy sub-band of polarized
electrons, the probability of capturing minority electrons is
small (see Fig.~\ref{BCapture}), whilst the probability of
capturing electrons from the lower-energy sub-band is large. Once
a cloud of approximately polarized electrons is captured, the
exchange interaction will generally entangle electrons in the same
dot together, so that the subsequent loss of electrons into the
Fermi sea will lead to a decoherent process that could relax the
remaining electrons into the low-energy polarized state. A more
detailed multiparticle analysis will be needed to determine the
final state of the remaining single electron, but the combination
of the above two processes is likely to lead to a high level of
polarization.

The above method is conceptually simple and would be easy to
implement in the laboratory. However, the macroscopic magnetic
field required in the capture region will need to be shielded from
the rest of the device where the quantum computation is to be
carried out, and this may present a nontrivial problem. If we
chose to drive spin flips using microwave pulses, then the
macroscopic field is actually required across the whole device.
However, the problem with using microwaves is their relatively
long wavelength which tends to affect every part the computation.
Alternatively, we may use local, static magnetic fields to
initialize, rotate and read out single qubits without a global
magnetic field. In Section~\ref{Section:Readout}, it will be
demonstrated that spin-polarized electrons can be prepared and
measured using a gate driven by the Stern-Gerlach effect. First we
turn to the implementation of single qubit rotations using local
static magnetic gates.

\section{Single-qubit unitary gates}\label{Section:Unitary}
Single-qubit unitary operations may be carried out, in principle,
by allowing the trapped single electrons to pass through regions
of uniform magnetic field. If there is no spin-orbit coupling
effect, the spin state of the qubit will evolve according to the
Zeeman term in the Hamiltonian: $\half g \BohrMag\,\bm{B}\cdot
\bm{\sigma}\,\psi$, where $\bm{B}$ is the magnetic field,
$\bm{\sigma}$ is the vector form of the 3 Pauli operators and
$\psi=(\alpha,\beta)$ is a spinor. The Bloch vector
$n=\left(-2\,\mathrm{Im}[\alpha\beta^*],2\,
\mathrm{Re}[\alpha\beta^*],|\alpha|^2-|\beta|^2\right)$ precesses
about the direction of the magnetic field with angular frequency
$g \BohrMag |B|/\hbar$. A local static magnet may be produced by a
magnetic force microscope or by evaporative deposition of a
ferromagnetic material such as Cobalt, or a permalloy such as
NiFe. Inevitably, different samples will produce different
strengths of magnetic field. But there are methods to vary the
strength and pattern of the field once the sample has been
fabricated. It has been demonstrated that ferromagnetic properties
of thin-film 3d transition metals can be modified via ion
irradiation.\cite{Kaminsky:2001} Another method would be to use
oxidation techniques with the atomic force
microscope.\cite{{Nemutudi:AFM:02},{Nemutudi:AFMLithog:01},{Curson:AFMQ1DC:01}}
Only two independent directions of the B-field are necessary to
produce an arbitrary single-qubit manipulation and we will choose
these to be perpendicular to the direction of the SAW, one aligned
with and the other perpendicular to the 2DEG.

For idealized qubits with no spatial degree of freedom, the above
model for single-qubit rotations is complete. However, the trapped
electron is a charged particle with a spatial distribution within
the dot. The fields couple to both the spatial and spin degrees of
freedom, causing the electron to experience the Lorentz force as
well as spin-precession. It is therefore clear that one cannot
increase magnetic fields arbitrarily, since the Lorentz force will
upset the confinement properties of the electron. Nor can the
direction of the field be chosen arbitrarily without consequences
for the robustness of the gate. To address these concerns, we
analyze the behavior of the full wavefunction: a two-component
spinor field defined over space and time, $\psi_\sigma(\bm{x},t)$,
under the action of gates operated by static magnets.

\subsection{Pauli Hamiltonian with uniform magnetic fields}
Assuming a uniform magnetic field in the lab frame, we solve the
Pauli equation for the spin field from which the probability
density field and the Bloch vector field can be obtained.

The qubit is trapped in a net electrostatic potential with
contributions from the Q1DC split gates, the 2DEG confining
potential and the SAW piezoelectric potential. The parameter
regimes we consider allow us to neglect motion out of the 2DEG
plane.\footnote{In other words, we ignore the $z$ dependence of
the wavefunction. This is partly justified if $e\,\beta\,\Delta
x/\sqrt{\Delta E\,m^*} \ll 1$, where $\Delta E$ is the
characteristic energy gap between the
 2DEG energy bands and $\Delta x$ is the
 length of the gate.
 This is indeed the case for parameter regimes that we consider.}
 The net confining potential in the $xy$ plane is modeled by
 \cite{Robinson:PRB:2001}
\begin{eqnarray}
 V &=& V_\mathrm{Q1DC} + V_\mathrm{SAW} \nonumber \\
   &=& V_0 \frac{y^2}{2 w^2}  + A\,
   \big\{1-\cos\left[2\pi(x/\lambda - f t)\right]\big\}.
   \label{V}
\end{eqnarray}

The Q1DC split gate voltages are such that typically $V_0 \sim
2800 \units{meV}$. The width of the Q1DC $w$ is typically between
$1$ and $2\units{\mu m}$. The amplitude of the SAW is $A \sim
40\units{meV}$ with wavelength $\lambda \sim 1\units{\mu m}$ and
frequency $f \sim 2.7\units{GHz}$.

The appropriate nonrelativistic equation for the two-component
spinor field is the Schr\"odinger equation with a Pauli
Hamiltonian. For an electron moving in an arbitrary vector
potential $A$ and potential energy $V$, the Pauli Hamiltonian $H$
is
\begin{equation}\label{PauliEq}
\frac{1}{2m^*} \left( \hat{p}^2 + e\,(\hat{p}\cdot A) +  2 e
A\cdot \hat{p} + e^2A^2\right) + \half\, g \BohrMag\, \sigma \cdot
B + V,
\end{equation}
where $e>0$ is the electronic charge, $g \simeq 0.44$ is the
Land\'e g-factor and $m^* \simeq 0.067 m_e$ is the effective mass
in GaAs. By simulating in the plane of the 2DEG we simplify to a
2+1 dimensional model, with variables $(x,y,t)$. If the particle
were chargeless but with an anomalous magnetic moment, terms in
$H$ that explicitly involve $e$ can be dropped. In such a case,
there is no Lorentz force acting on the particle, the
Schr\"odinger equation simplifies significantly and there would be
no need to concern ourselves with the spatial behavior of the
qubit.

Before proceeding further, it is convenient to transform into the
rest frame of the electron moving with the SAW speed $v$. We may
use simple Galilean transformations $ x' = x - vt,\ y' = y,\ t' =
t$, since the SAW velocity is non-relativistic. The potentials in
the electron rest frame become, in the harmonic oscillator
approximation,
\begin{eqnarray}
 A(x',t')  &=& -\beta(x'+ v t') \\
 V(x',y') &=& V_0 \frac{{y'}^2}{2 w^2}+ A \left[1-\cos\left(
 \frac{2\pi x'}{\lambda}\right)\right]\\
&\simeq& \frac{V_0}{2 w^2}\, {y'}^2 + \frac{A}{2}\, k^2 {x'}^2,
 \label{V'}
\end{eqnarray}
with $k=2\pi/\lambda$. A further convenience is to use a system of
natural units such that $\hbar,m^*,v,e$ are unity. The units of
length, time and energy become $\hbar/m^*v=0.640\units{\mu m}$,
$\hbar/m^*v^2=0.237\,\mathrm{ns}$ and $m^*v^2=2.78\,\mathrm{\mu
eV}$, respectively. The natural unit of magnetic field is $1.61
\units{mT}$. Parameters can now be assigned with dimensionless
values with respect to the above units. In the electron rest
frame,
\begin{equation}\label{PauliEq2}
\begin{split}
 -\half \nabla^2 \psi -
i\left[\,A\cdot\nabla+\half(\nabla\cdot A)\,\right]
 \psi + \half A^2 \psi \\
 + \half\, g \BohrMag
\, \sigma\cdot B \psi + V \psi = i \pderiv{}{t}\psi,
\end{split}
\end{equation}
where the primes will subsequently be dropped from the
coordinates. We will further simplify the Hamiltonian by adopting
the Coulomb gauge $\nabla\cdot A=0$.

\subsection{Qubit rotation: Uniform transverse magnetic field}
\begin{figure}[!t]
 \centering
 \includegraphics[width=8.4cm]{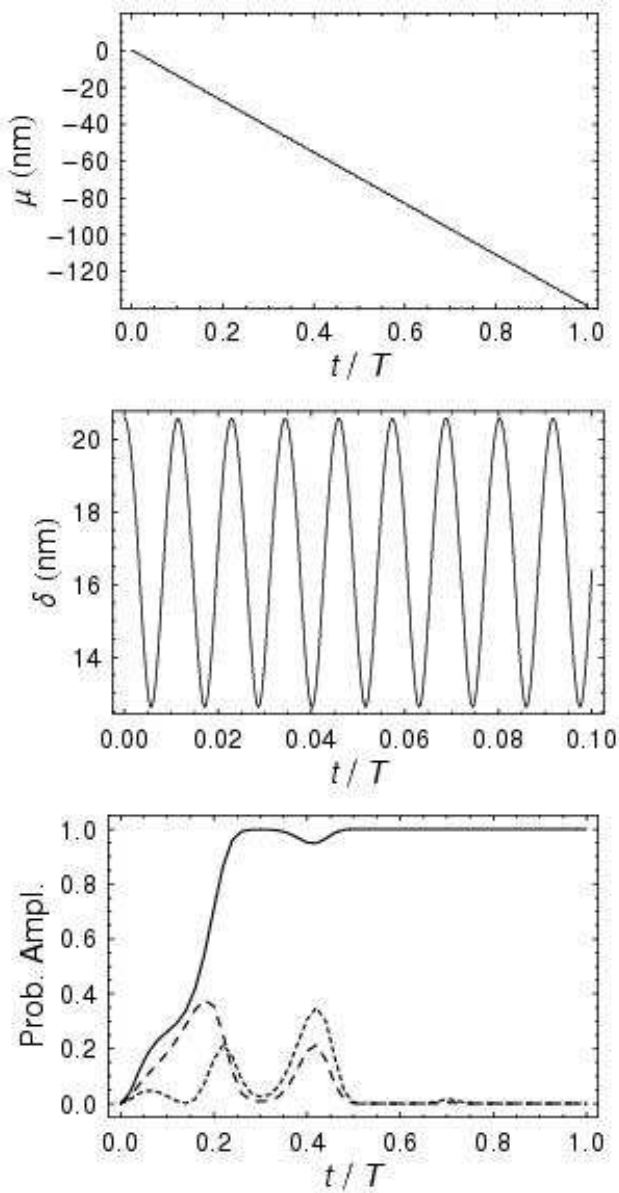}
 \caption{
 Gaussian evolution of the probability
  distribution $|\chi(x,t)|^2$
 in a constant magnetic field $B_y$. The top plot shows
  peak position $\mu$ and the middle plot shows the width $\delta$.
The bottom plot shows the probability amplitudes $|C_n|^2$ of the
$n$th SHO mode in the $x$ direction: $|C_1|^2$ (long dash),
$|C_2|^2$ (short dash). The plot also shows $1-|C_0|^2$ (solid).
 Typical values for the SAW dot are taken:
 $A = 40\units{meV}$; $\lambda = 1.0\units{\mu m}$; magnetic
 field of $1\units{T}$. $T= \pi/g\BohrMag\beta\approx 0.08\units{ns}$
 is the time taken for a $\pi$ rotation of the Bloch
 vector about the $y$-axis. The initial Gaussian wavefunction was
 taken to be the
 SAW quantum dot groundstate $(C_0=1)$.
 In the limit $c_1/c_0 \ll 1$ (weak field), $\mu(t)$ moves
 approximately linearly and $\delta(t)$ undergoes bounded
 oscillations about $\delta(0)=s/\sqrt{2}$.
 \label{GaussianSolution}
 }
\end{figure}

For a magnetic field in the $y$ direction in the plane of the 2DEG
and transverse to the Q1DC, we model the magnetic field in the
region of the single-qubit gate with a vector potential of the
form
\begin{equation}\label{Az}
  A_z(x) = -\beta\,x,
\end{equation}
with $A_x=A_y=0$. Clearly this does not vanish at infinity, but we
only consider interactions over regions of finite extent. This
potential generates a uniform magnetic field of strength $\beta$
in the $y$ direction. With this potential Eq.~(\ref{PauliEq2})
becomes
\begin{equation}\label{PauliEq3}
 -\half \nabla^2 \psi + \half\, \beta^2(x+t)^2 \psi + \half\, g \BohrMag
 \beta\,\sigma_y\psi + V\psi = i\pderiv{}{t}\psi,
\end{equation}
in which the potential $V$ is given by (\ref{V'}). An effective
potential $V + A^2/2$ can be identified, which resembles a
harmonic oscillator but is time-dependent. Anticipating an
analytic solution by separation of variables we apply the ansatz
\begin{equation}
 \psi_\pm(x,y,t) =
 \chi(x,t)\,\phi_n(y)\,e^{-iE_nt}|s_y\rangle\,e^{- i s_y \Delta E\,
 t/2},
 \label{ansatz}
\end{equation}
where $\phi_n$ are the harmonic oscillator eigenstates with
energies $E_n=\omega_y (n+\half)$, and $|s_y\rangle$ are the
eigenstates of $\sigma_y$ with eigenvalues $s_y=\pm 1$. We have
introduced the oscillator frequency $\omega_y = \sqrt{V_0}/w$ and
the Zeeman energy gap $\Delta E = g \BohrMag\beta$. The energy
eigenstates in the $y$ direction are exactly the harmonic
oscillator modes. The wavefunction in the $x$ direction is
time-dependent and it is the solution of $\chi(x,t)$ to which we
now turn. On substituting (\ref{ansatz}) into (\ref{PauliEq3}), we
obtain a PDE for $\chi(x,t)$:
\begin{equation}\label{chiEquation}
 -\half\, \partial^2_x\, \chi + \left[\half(c_0+c_1)\,x^2 +  c_1\,xt\right]\chi = i\, \partial_t\, \chi,
\end{equation}
with $c_0= A\,k^2$ and $c_1= \beta^2$. Terms which depend only on
$t$ have been dropped, as they merely contribute global,
time-dependent phases that do not affect the dynamics. A Gaussian
solution of (\ref{chiEquation}) can be found with a further ansatz
\begin{equation}\label{GaussAnsatz}
 \chi(x,t) = \exp\left[f_1(t)\,x^2 + f_2(t)\, x + f_3(t)\right].
\end{equation}
This system of time-dependent functions can be determined
self-consistently, assuming a Gaussian groundstate of the SAW dot
for the initial condition at $t=0$. The resulting expressions for
$f_1,f_2,f_3$ are complicated, but by noting a few general
properties of the solution we can understand all the important
features of the dynamics. $f_3(t)$ takes account of the
normalization but is otherwise of no more interest. $f_1(t)$ and
$f_2(t)$ together describe a groundstate Gaussian wavefunction
evolving in the time-dependent vector potential. The solution is
particularly simple in that it remains Gaussian throughout, so we
will only need to keep track of the position of the central peak
$\mu$ and the width (standard deviation) $\delta$ of the
probability distribution $|\chi(x,t)|^2$:
\begin{eqnarray}
 \mu(t) &=& -\half \frac{\mathrm{Re}[f_2(t)]}{\mathrm{Re}[f_1(t)]} \\
 \delta(t) &=& \half\, \sqrt{-\mathrm{Re}[f_1(t)]}.
\end{eqnarray}
The typical energy scales for the SAW electron encountered in
experiment puts the system in the regime where $s \ll 1$ and $c_1
\ll c_0$. Expanding $f_1$ and $f_2$ to first order in $s$ and
$c_1/c_0$, we obtain the asymptotic behavior for the position of
the peak:
\begin{equation}
 \mu(t)\rightarrow \frac{-c_1}{c_0+c_1}\,t.
\end{equation}
From this result we see that the magnetic field introduces a
constant drift velocity of the peak in the $-x$ direction. On
exiting the interaction region, the peak will be off-center with
respect to the SAW dot and the probability distribution will
subsequently oscillate in the $x$ direction, perhaps exciting
higher-energy orbital states of the dot. If the charge
distribution is pulled too far off-center, it is likely to escape
the SAW quantum dot. We could ask how long the electron can remain
in the field before it is dragged a distance $\lambda$ away from
the center of the dot in the $x$ direction:
\begin{equation}
 T_\mathrm{max} \simeq \frac{c_0+c_1}{2 c_1}\lambda.
\end{equation}
When this is compared with the time required for the Bloch vector
to rotate by some appreciable angle such as a $\pi$ rotation, we
obtain the ratio
\begin{equation}
T_\mathrm{max}/T_\pi= g\BohrMag
\left(\frac{A\,k^2+\beta^2}{k}\right),
\end{equation}
which is about $2000$ for a $1\units{T}$ field. This means that
the Bloch vector is allowed to precess approximately $1000$ times
before the drag effect destabilizes the qubit confinement. In the
limit of weak fields, this ratio is no smaller than
$g\BohrMag\,A\,k\sim 800$. Therefore, whilst it is possible that
higher orbital states are excited as a consequence of the
interaction, there is little danger of the electron escaping the
SAW quantum dot during the operation of the gate. We now turn to
the behavior of $\delta(t)$, the width of the Gaussian
wavefunction, which in the limit $c_1 \ll c_0$ oscillates
according to the simple expression
\begin{equation}
 \delta(t) =  \sqrt{\frac{\sqrt{c_0}}{2(c_0+c_1)}}
 \left[1+\frac{c_1}{c_0}\,\cos^2(\sqrt{c_0+c_1}\ t)\right]^{1/2},
\end{equation}
with $c_0 = A\,k^2$ and $c_1 = \beta^2$. The frequency of the
oscillation increases with the energy of the dot and is typically
much faster than the Zeeman spin precession frequency, i.e., rate
of precession of the qubit Bloch vector. These results are plotted
for a specific case in Fig.~\ref{GaussianSolution}. The top two
plots show the evolution of $\mu$ and $\delta$. As the Bloch
vector rotates about the field along the $y$-axis, the Gaussian
translates in the $x$ direction with a rapidly oscillating width.
The bottom plot of Fig.~\ref{GaussianSolution} shows probability
amplitudes of excitation into higher-energy simple harmonic
oscillator (SHO) modes in the $x$ direction of the quantum dot.

Our conclusions are as follows. In addition to rotating the Bloch
vector as required, the gate has the effect of displacing the
center of the Gaussian wavefunction which will increase the energy
of the bound state by an amount depending on both the gate time
and the field strength. Higher-energy orbital states are likely to
be excited, which could lead to decoherence via spin-orbit
couplings or dipole coupling to phonons and photons. According to
Fig.~\ref{GaussianSolution}, the electron is almost completely out
of the groundstate after a $\pi/2$ rotation of the Bloch vector.
In the extreme case after many spin precessions (typically
thousands), the qubit will leave the SAW quantum dot. However, for
quantum computations the gate time need only be long enough to
conduct a single orbit of the Bloch sphere, in which case the
excitation into higher orbital states is negligible. These
observations reveal some important features of the gate operation
that are not revealed by an idealized spin-only model of the
qubit.

\subsection{Qubit rotation: Uniform perpendicular magnetic field}
To move the qubit state to an arbitrary point on the Bloch sphere
a second axis of rotation on the Bloch sphere is needed, and to
this end we consider a gate implemented by a uniform magnetic
field in the $z$ direction. This will not be just a trivial
extension of the preceding analysis, since the 3D rotation
symmetry is broken by the motion of the SAW in the $x$ direction.
In practice, the $y$ magnetic field may be easier to fabricate
than the $z$ magnetic field, since the latter passes
perpendicularly through the 2DEG structure. It could feasibly be
produced by layering oppositely aligned thin-film magnets just
beneath and just above the 2DEG, or alternatively by applying a
global $z$ magnetic field which is shielded in regions where it is
not needed.

A vector potential generating the uniform $B_z$ magnetic field is
\begin{equation}\label{Ax}
 A_x(y)=-\beta y,
\end{equation}
with $A_y=A_z=0$. It should be noted that although the field is
uniform and static in the laboratory frame, the electron sees a
moving uniform field. In the electron rest frame, $A_x$ is
time-independent and $z$-independent, allowing a Hamiltonian in
$x$ and $y$ only. The chosen gauge facilitates numerical
simulations which follow shortly. An effective scalar potential
that is quadratic in $y$ arises from the $A^2$ term in the
Hamiltonian (\ref{PauliEq2}). It is interesting to compare the
strength of this confining potential with the Q1DC potential,
which is also approximately quadratic in y:
\begin{equation}\label{CompareEnergies}
 \frac{\textrm{Energy of $A^2$ term}}{\textrm{Energy of Q1DC
 term}} = \frac{\beta^2}{V_0/w^2}.
\end{equation}
With magnetic fields of order 1 Tesla and typical Q1DC energies,
this ratio is of the order unity. This means that for the
parameter values being considered the effective scalar potential
arising form the $A^2$ term is comparable to the Q1DC confinement
potential. A stronger field would start to significantly deform
the shape of the dot. This will not be a problem as long as the
evolution has occurred adiabatically during the deformation.
However, as we shall discuss in a moment, the probability of
excitation into higher orbital states due to the $B_z$ field is
not negligible.

The other interesting term in the Hamiltonian is the asymmetric
coupling $ -i\beta y\,
\partial_x$ between the $x$ and $y$ variables. This is expected to
introduce rotational behavior, as we would intuitively anticipate
some form of Landau orbital motion due to the Lorentz force.

\begin{figure}
 \centering
 \includegraphics[width=3.3in]{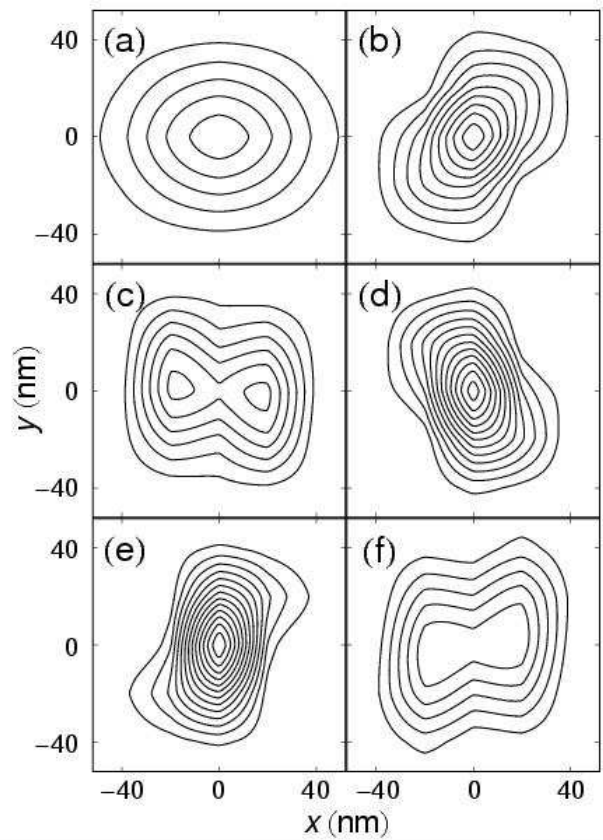}
 \caption{
Contour plots of electron probability density during the operation
of a qubit rotation gate. This is a numerical simulation in the
$xy$ (2DEG) plane of a SAW electron
  undergoing a $\pi/2$ rotation gate about the $z$ axis, under a
uniform  magnetic field $B_z$ of strength $1\units{T}$. Snap-shots
are shown at the following times (ps): (a) 0, (b) 2.89, (c) 11.6,
(d) 20.3, (e) 28.9, (f) 37.6. The initial probability distribution
is a Gaussian groundstate with standard widths $\sigma_x = 26.0$
and $\sigma_y = 20.6$.
 The probability density rotates about the
$z$
  axis under the influence of the Lorentz force acting
 in the $xy$ plane. Moreover, an effective scalar potential
 contributed by the $A^2$ term in the
 Hamiltonian is quadratic in $y$, which spatially
 squeezes the initial Gaussian distribution into
 the center of the dot.
 \label{rotation:z}
 }
\end{figure}

A numerical simulation implementing a Crank-Nicholson,
\cite{CrankNicolson} finite difference algorithm (alternating
direction method) was used to simulate the operation of the gate.
We started with an initial Gaussian groundstate of the dot with
spin state $|\!\uparrow_x\rangle$ and subjected it to $1\units{T}$
of magnetic field for a duration of $T_{\pi/2} = \pi/2 g \BohrMag
\beta$, which is the gate time required for a $\pi/2$ rotation of
the Bloch vector about the $z$ axis. The evolution of the Bloch
vector is simple due to the uniformity of the magnetic field:
$n_x(t)=\cos(g\BohrMag\beta t)$, $n_y(t)=\sin(g\BohrMag\beta t)$,
and $n_z(t)=0$. All other parameters were assigned those values
given just after Eq.~(\ref{V}). The result of the simulation is
shown in Fig. \ref{rotation:z}, which shows time-shots of the
probability distribution in the 2DEG ($xy$) plane. Losing its
initial elliptic contours, the distribution develops two lobes
which rotate about its midpoint. The density at the center
increases due to the spatial squeezing from the $A^2$ term in the
Hamiltonian. This clearly shows excitation into higher-energy
orbital states.

Using perturbation theory to second order in the strength of the
field $\beta$, with harmonic oscillator modes as the basis, we
found the following facts: (i) To second order of perturbation
theory, only the second excited state with SHO quantum numbers
$n_x=n_y=1$ becomes populated; (ii) The probability ratio with
respect to the groundstate is
\begin{equation}\label{OscPert}
\left|\frac{C_{11}}{C_{00}}\right|^2 = {\beta }^2\, \frac{\omega_x
\sin^2[t\,(\omega_x + \omega_y)/2]}{\omega_y( {\omega_x}
+{\omega_y})^{2}},
\end{equation}
where $\omega_x$ and $\omega_y$ are frequencies arising from the
harmonic oscillator approximation; (iii) The amplitude of the
ratio approaches unity when the field approaches $1\units{T}$;
(iv) The ratio of the spin precession frequency to the frequency
of the $|C_{11}|$ oscillation is about $1.6\times 10^{-5}$ for
typical SAW parameters (see just after Eq.~(\ref{V})).

The above model is useful in assessing the robustness of the qubit
and its susceptibility to decoherence due to orbital motion.
Population into this excited state is a problem for decoherence,
since the oscillating charge in the dot couples via dipole
interactions to phonons and other charges outside the dot.
However, provided we have sufficient control of the gate time, we
can use (\ref{OscPert}) to make the electron exit from the gate in
its groundstate. Otherwise, a gate driven by radiofrequency pulses
in the presence of a global magnetic field could provide an
alternative means to implement spin rotations about the $z$ axis.

\section{Single-qubit initialization and measurements}\label{Section:Readout}
In addition to single-qubit rotation gates, we require the ability
to initialize and measure qubits at the beginning and end of the
computation. Spin-polarized electrons can be obtained from
injection through a ferromagnetic
contact.\cite{Datta:APL:1990,Ohno:Nat:1999} There is also a method
to polarize spin using nondispersive phases (Aharonov-Bohm and
Rashba) without the need for ferromagnetic
contacts.\cite{Ionicioiu:PRB:2003} In the field of quantum
computing, a well-known method for achieving readout of solid
state spin qubits is to convert spin information into charge
information\cite{Loss:PRA:1998} and subsequently use
single-electron transistors or point contacts to detect charge
displacements or Rabi
oscillations.\cite{Pakes:2003,{Friesen:2003}} However, recent
theoretical
 results by Stace and
Barrett \cite{Stace:PRL:2004} argue the absence of coherent
oscillations in a continuously measured current noise, contrary to
previous results and
assumptions,\cite{Goan:PRB:2001,Korotkov:PRB:2001} and therefore
raise concerns about the measurability of charge oscillations in
similar scenarios. In any case, it is difficult to apply these
methods to qubits in moving quantum dots. We therefore turn
towards a quite different approach, one which enables the
initialization and readout of SAW electron spin qubits solely with
the aid of nanomagnets and ohmic contacts.\cite{Barnes:PRB:2000}
The readout gate we consider is based on the Stern-Gerlach
effect.\cite{Stern-Gerlach:1922,Challinor:96,{Venugopalan:DecohSternGerlach:95}}
In the 1920s Bohr and Pauli asserted that a Stern-Gerlach
measurement on free electrons was
impossible,\cite{bohr'sArg,Pauli:6thSolvay} using arguments which
combined the concept of classical trajectories and the uncertainty
principle. This subsequently led physicists to analyze
single-electron Stern-Gerlach measurements within increasingly
more rigorous quantum settings, ultimately ending the debate by
showing that the measurement can indeed be done, albeit with
certain caveats.\cite{Batelaan:97,{Garraway:99},{Gallup:01}} Thus
Stern-Gerlach measurements on \emph{free} electrons have been
extensively investigated, but little attention has been given to
such measurements on confined electrons.\cite{dehmelt2} An
interesting semiclassical analysis of a Stern-Gerlach type
experiment with conduction electrons has been
reported,\cite{Fabian:2002} in which the authors justifiably
neglect the Lorentz force effects. In contrast, we analyze a
single-electron Stern-Gerlach device, providing a full quantum
mechanical treatment and including all Lorentz force effects.

In the SAW electron system, single electrons are transported in
moving quantum dots. Can this system undergo a Stern-Gerlach type
measurement, and would it be of any use in a scheme for quantum
computation using qubits trapped in surface acoustic waves?

\begin{figure}
 \centering
 \includegraphics[width=8.6cm]{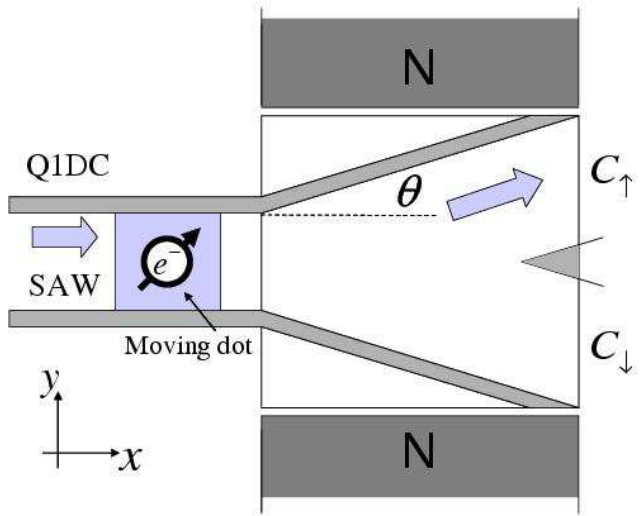}
 \caption{
 Schematic diagram of the spin
 readout/polarizing device based on the Stern-Gerlach effect: SAW
 propagates from left to right transporting a single electron
 in a moving quantum dot. Q1DC relaxes with a
gradient $\tan(\theta)$ to partially delocalize the particle
during the gate operation; $C_\uparrow$($C_\downarrow$) labels the
Q1DC receiving electron flux in the spin up(down) state of the
$\sigma_y$ operator; Magnets (of any geometry and polarity) need
to produce a localized, inhomogeneous distribution of magnetic
field.
 \label{SGDevice}
 }
\end{figure}

The electron confinement to low dimensions allows us to guide the
electron through the magnetic field in ways that enhance the spin
measurement and suppress the deleterious effects of delocalization
and the Lorentz force. Surface magnets can be arranged in such a
way as to produce a local magnetic field inhomogeneity. For
example, two north poles placed on either side of the Q1DC will
produce a region of intense magnetic field gradient inbetween (see
Fig.~\ref{SGDevice}). Via the Zeeman interaction term
$\propto\sigma\cdot B$ in the Pauli Hamiltonian, the field
inhomogeneity has the effect of correlating the spatial location
of a wavepacket to its spin state.\cite{Peres:QTCM}
In most situations it is  necessary to continue confining the
qubit during the operation of the gate, because the spreading
time\footnote{The spreading time for a Gaussian state is the time
taken for its standard deviation to double.} for a free Gaussian
wavefunction is comparatively short -- on the order of
$1\,\mathrm{ns}$. But the gate must cause a wavepacket splitting
in order for the spin states to be resolved, hence the Q1DC must
relax to allow for motion in the $y$ direction. This is achieved
by patterning the Q1DC in a funnel shape, with an opening angle
$\theta$ (see Fig.~\ref{SGDevice}), such that in the electron rest
frame the potential looks like
\begin{equation}\label{VQ1DC}
 V_\textrm{Q1DC} = V_0\, y^2 /
 2\left[w+\tan(\theta)u(t)t\right]^2,
\end{equation}
where $u(t)$ is the step function: $u(t<0)=0$ and $u(t\ge 0)=1$. A
negatively biased surface gate placed on the $x$ axis can be used
to guide the electron into the channels. If necessary, both the
position of the electrode and the opening angle $\theta$ could be
optimized for a particular sample device under low-temperature and
high-vacuum conditions by erasable electrostatic
lithography.\cite{Cook:2003} We will study the quality of the
readout obtained from the gate for two different magnetic field
configurations, both sufficiently simple so as to be realizable in
the near future: the linearly inhomogeneous field and the 2D
dipole field.

\subsection{Stern-Gerlach gate using a linearly inhomogeneous field}
In the first model we will analyze a simple, unidirectional and
linearly inhomogeneous field pointing in the $y$ direction:
\begin{equation}\label{UniformInhomo}
 B_y = - \beta\, y,
\end{equation}
with $B_x=B_z=0$. A wedge-shaped single domain surface magnet of
appropriate dimensions can produce an inhomogeneous magnetic field
of this form near its center of symmetry. A vector potential for
this field is $A_x=-\beta\, y z$, with $A_y=A_z=0$. This field
exerts a spin-dependent force in the $y$ direction and it is the
simplest field that induces the Stern-Gerlach effect. Although a
$z$ dependence enters into the vector potential, by considering
motion only in the $z=0$ plane we may avoid contributions from
terms involving $A$ in the Hamiltonian, as well as the $z$
component of magnetic field. The absence of $x$ in the potential
immediately allows us to write down harmonic oscillator modes for
the $x$ dependence. The remaining $(y,t)$ dependent part obeys
\begin{equation}\label{diffeq}
 \left(
 -\half \nabla^2  - \half\, g \BohrMag  \beta s_y\, y
  + V_\mathrm{Q1DC}(y,t)- i\,\partial_t \right) \psi(y,t) = 0
\end{equation}
where $s_y$ is the eigenvalue $\pm 1$ of the spinor $|s_y\rangle$,
and $V_\mathrm{Q1DC}$ is given by (\ref{VQ1DC}).  The initial
state is again the Gaussian ground state of the Q1DC. A solution
to (\ref{diffeq}) is obtained by a time-dependent Gaussian ansatz
(\ref{GaussAnsatz})
\begin{equation}
 \psi(y,t) = \exp\left[f_1(t)\,y^2 + f_2(t)\, y + f_3(t)\right].
\end{equation}
In a similar way as before, we solve the system of coupled
ordinary differential equations and derive the time dependence of
the standard deviation $\delta(t)$ and the position $\mu(t)$ of
the probability distribution.

Let $c_0=V_0/w^2$ and $c_1=g \BohrMag \beta s_y$. The width of the
initial Gaussian groundstate wavefunction is determined by $s=
c_0^{-1/4}$. The behavior of the Gaussian solution is then
characterized by the following time-dependent parameters
\begin{eqnarray}
\delta(\tau)^2  &=&  \frac{\tau }{2\,\gamma\,\sqrt{c_0}} \,
     \big[
      -4\,c_0 +
      {\alpha }^2\,\cosh
      (\sqrt{\gamma}\,\ln(\tau)/\alpha)\nonumber \\
     &&-
      \alpha \,{\sqrt{\gamma }}\,\sinh
(\sqrt{\gamma}\,\ln(\tau)/\alpha)
     \big]   \\
 \mu(\tau) &=& \frac{\sqrt{\tau}\,c_1}{2\,(2\,\alpha^2+c_0)}
           \big[
                \tau^{3/2} -
                        \cosh
(\sqrt{\gamma}\,\ln(\tau)/\alpha) \nonumber\\
               &&-
               \frac{3\alpha}{\sqrt{\gamma}}
                 \sinh
({\sqrt{\gamma}\,\ln(\tau)/\alpha}) \big],
               \label{MuEvolution}
\end{eqnarray}
where we have introduced ancillary variables $\alpha=
\tan(\theta)/w$, $\gamma = \alpha^2 - 4 \,c_0$ and $\tau = 1+
\alpha \,t$. The parameter $\gamma$ is useful in determining when
the trigonometric functions become oscillatory. If the angle
$\theta$ is critical such that $\tan(\theta)=2\sqrt{V_0}$, then
$\gamma=0$ and the width has an especially simple behavior:
\begin{equation}
\delta(t)^2= \frac{ \tau }{\alpha}
 \,\left[
  1
   - \ln\tau
  + \half\,(\ln\tau)^2
  \right].
\end{equation}
This model predicts the width and position of the qubit
wavefunction after a time $t$ in the linearly inhomogeneous field
with field gradient $\beta$. In practice we cannot produce
arbitrary magnetic fields in arbitrary configurations. For a weak
magnetic field such that $c_0 \gg c_1 \gg \alpha$,
\begin{eqnarray}
 \delta(\tau) &=& \sqrt{\frac{\tau}{2 \sqrt{c_0}}}
  =
 \delta(0)\,\sqrt{ 1 + \tan(\theta)\, \frac{t}{w}
 } \\
 \mu(\tau) &=& \frac{c_1}{2\,c_0}
   \left[
    \tau^2 -
    \sqrt{\tau}\,\cos\left(\frac{\sqrt{c_0}}{\alpha}\ln\,\tau\right)
   \right].
\end{eqnarray}
The width increases as $\sim\sqrt{t}$ and the position moves as
$\sim t^2$ as expected. We may compare $\delta(t)$ with the
dispersion of a Gaussian in free space
\begin{equation}
 \delta(t) = \sqrt{\delta(0)^2 + \frac{t^2}{4\,\delta(0)^2}}\ ,
\end{equation}
which is linear in $t$ at large times. The broadening of the width
is suppressed due to the confinement potentials. The effect of the
Stern-Gerlach can be pictured as follows. Consider a qubit in the
state $|\!\uparrow_y\rangle$ entering the gate. The Q1DC broadens
as the qubit enters the field, allowing the Stern-Gerlach effect
to produce a spin-dependent shift in the center of mass towards
the channel $C_\uparrow$. However, the initial localized
distribution will deloclize due to the Q1DC broadening, allowing a
small current to be detected in the wrong channel $C_\downarrow$.
Thus this gate is a spin polarizing filter with some intrinsic
error rate, independent of decoherence effects, shot noise and
Johnson noise. It remains for us to judge how well it performs,
and to this end we introduce a fidelity measure of the gate.

\subsection{Fidelity of the Stern-Gerlach
gate with linearly inhomogeneous field} 
\begin{figure}
 \centering
 \includegraphics[width=3.4in]{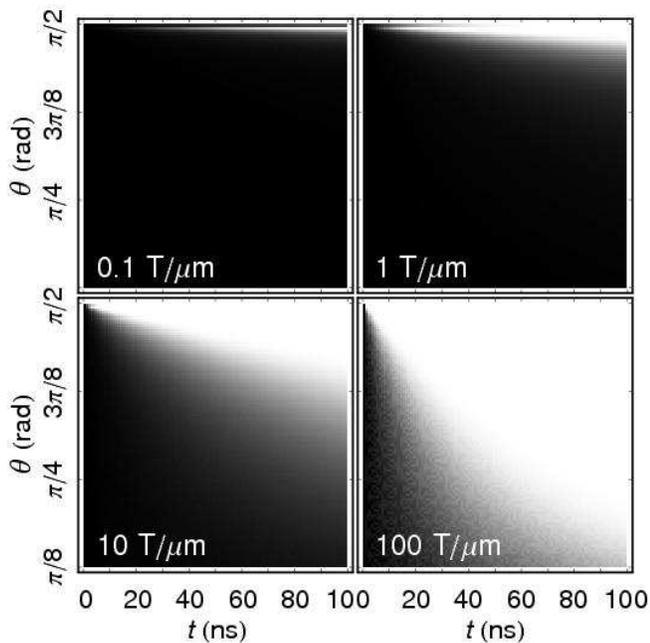}
 \caption{
 Fidelity $F_\uparrow$
  in the $C_\uparrow$ channel for a
 Stern-Gerlach gate operated
 by a linearly inhomogeneous and unidirectional magnetic field
  in the $y$ direction. White regions $\Rightarrow$
  $F_\uparrow=1$.
 $\theta$ is the angle between the Q1DC and SAW direction.
 $t$ is the gate time. Input state
 is a maximally-mixed spin state in the
 groundstate of the SAW quantum dot.
 Various magnetic field gradients have been chosen to show
 the increase in the optimal parameter region (white)
 in the $(t,\theta)$ plane with increasing field
 gradient. $t_\mathrm{max}\approx 100\units{ns}$ is a crude
 order of magnitude estimate for the GaAs spin relaxation lifetime.
 \label{fidelitycontour}
 }
\end{figure}

As an input to the
Stern-Gerlach gate, we consider a density matrix of the form
\begin{equation}
 \rho(y,y') = g(y)\,g(y')^* \otimes \sigma,
\end{equation}
where $g(y)$ is the Gaussian groundstate of the dot and $\sigma$
is the initial spin density matrix.  In the inhomogeneous field
the spatial degrees of freedom couple to the spin degrees of
freedom, leading to
\begin{equation}
 \rho(y,y',t) = \sum_{s,s'=\uparrow,\downarrow}
 g_s(y,t)\,g_{s'}(y',t)^* \, \sigma_{s\,s'}|s\rangle\langle s'|.
\end{equation}
After a gate time $t$, we post-select the state in, say, the
$C_\uparrow$ channel and renormalize:
\begin{eqnarray}
 \tilde{\rho}(y,y',t) =
 \frac{1}{N(t)}\, \sum_{s,s'} \sigma_{s\,s'} |s\rangle\langle
 s'| \times \nonumber\\
   \int dz\,dz'\,
   E_\uparrow(y,z)\,g_s(z,t)\,g_{s'}(z',t)^*\,E_\uparrow(z',y'),
\end{eqnarray}
The probability yield in the channel is the
normalization  $N(t)$ of the post-selected state. The simplest
projection kernel projecting into the support  of $C_\uparrow$ is
$E_\uparrow(y,y') = \delta(y-y')\,u(y)$, where $u(y)$ is the
step-function defined in (\ref{VQ1DC}). We finally define our
fidelity as
\begin{eqnarray}
 F_\uparrow  &=& \int dy\, \langle \uparrow | \tilde{\rho}(y,y,t) |
 \uparrow\rangle \\
            &=& \frac{ \sigma_{ \uparrow \uparrow} }{N(t)}\int_0^\infty
            |g_\uparrow(y,t)|^2 dy.
\end{eqnarray}
The yield $N(t)$ can be calculated without explicit knowledge of
$g_\downarrow$, since the Gaussian solutions are related by
reflection symmetry: $g_\uparrow(y,t)=g_\downarrow(-y,t)$. We
choose various magnetic field gradients ranging from
$0.1\units{T\,\mu m^{-1}}$ to $100\units{T\,\mu m^{-1}}$ and plot
the output fidelity of a maximally-mixed input state as a function
of the Q1DC angle $\theta$ and the gate time $t$ (see Fig.
\ref{fidelitycontour}). Perfect spin filtering corresponds to a
fidelity $F=1$, and the worst-case fidelity is $F=1/2$. The
density plots show how the optimal parameter regions increase with
larger magnetic field gradients. This model is useful in
determining the appropriate gate geometry and field gradients
required to achieve good fidelities.

The technology of nanoscale single domain magnets is not yet
capable of producing large fields in arbitrary configurations. The
above simple linearly inhomogeneous magnetic field is therefore a
simple approximation to the real magnetic field configuration on
any particular device. In the next subsection we consider the
effect of a different (perhaps more realistic) field
configuration, namely, the field near one of the poles of a dipole
magnet.

\subsection{Stern-Gerlach gate with dipole
field}\label{DipoleField} In this subsection we study the quality
of the readout obtained from a Stern-Gerlach type gate driven by a
simple 2D dipole field lying in the 2DEG plane, as shown in
Fig.~\ref{VectorField}. In Appendix~\ref{DeriveDipoleField}, we
derive the magnetic vector potential
\begin{equation}\label{potential1}
A_z =  \frac{\beta x}{x^2 +(y-d)^2},
\end{equation}
where $\beta$ is a strength parameter and $d$ is the distance of
the 2D dipole from the Q1DC.

\begin{figure}
 \centering
 \includegraphics[width=3.6in]{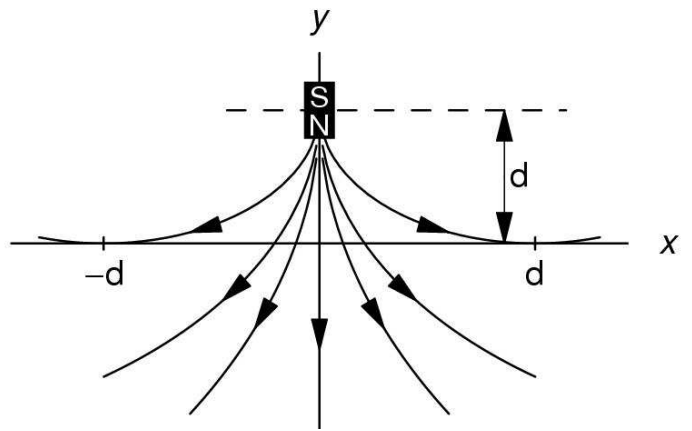}
 \caption{
 Vector field plot in the $xy$ (2DEG) plane showing the magnetic
 field due to an infinite string of
 dipole moments running parallel to the  $z$-axis at a
  distance $d$ from the Q1DC.
 \label{VectorField}
 }
\end{figure}

Figure~\ref{VectorField} shows that the gradient of the field in
the $y$ direction reverses direction at two points, $x=\pm d$. In
order to prevent the moving qubit from experiencing opposite field
gradients, we will restrict the gate time to $T=2\,d$ so that the
qubit moves in the region $ -T/2<x<T/2$ in the laboratory frame.
The question of how we are to shield the extraneous field from the
qubit does not have a simple solution. It has been suggested that
a superconducting material such as Niobium could be used to shield
the magnetic field where it is not needed.\cite{Barnes:PRB:2000}

\begin{figure}
 \centering
 \includegraphics[width=8.6cm]{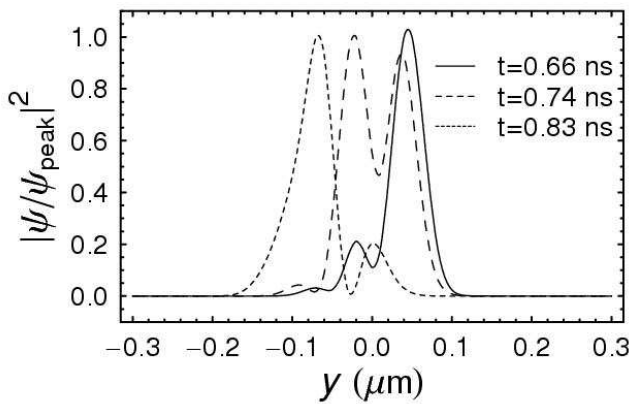}
 \caption{
 Three time-shots (see legend) showing the spatial electron
 probability density (normalized to peak). $y$ is the direction
 transverse to the Q1DC. This is a Stern-Gerlach type readout gate
 using the 2D dipole field ($1\units{T\mu m^{-1}}$ at $x=y=0$)
 shown in Fig.~\ref{VectorField}.
 Initial condition is a Gaussian wavepacket
 with spin $|\!\uparrow_\mathrm{y}\rangle$.
 Sideways translation of the probability density
  is suppressed due to both the strongly confining
  effective potential arising
 from the $A^2$ term and the precessional motion of the Bloch
 vector in the non-unidirectional field.
 \label{DipoleFieldSimulation}
 }
\end{figure}

The results of a 1D simulation of the wavefunction along the
$x=z=0$ direction are shown in Fig.~\ref{DipoleFieldSimulation}.
The numerical method used was a Crank-Nicholson algorithm adapted
to the Pauli Equation. The plot shows three time-shots of the
probability density $|\psi(y)^2|$, normalized to peak. The initial
state is a Gaussian groundstate with spin state
$|\!\uparrow_y\rangle$. The simulation shows no significant
spatial displacement of the electron density in the $y$
directions, in contrast to the behavior exhibited under the
linearly inhomogeneous field (c.f.~Eq.~(\ref{MuEvolution})). This
is partially due to the effect of the Lorentz force generated by
the $A^2$ term in the Hamiltonian, which contributes an effective
potential that is highly confining in the $y$ direction. The $A^2$
confinement is so strong that the Q1DC opening angle $\theta$
becomes irrelevant. Bohr and Pauli's claim that the Lorentz force
washes out the Stern-Gerlach effect are upheld in this particular
scenario.

\begin{figure}
 \centering
  \includegraphics[width=8cm]{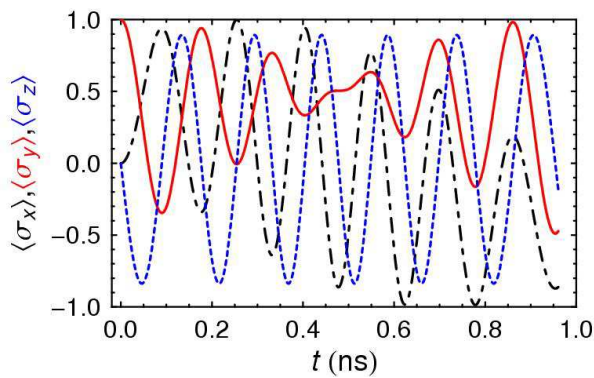}
 \caption{
 (Color online)
 Evolution of the Bloch vector in a 2D dipole field
 ($1\units{T\mu m^{-1}}$ at $x=y=0$): $\langle\sigma_x\rangle$
 (long-short dash), $\langle\sigma_y\rangle$ (solid),
 $\langle\sigma_z\rangle$ (short dash).
 Time $t=0$ corresponds to the starting point $x=-d$
  (see Fig.~\ref{VectorField}).
 We have used a
spin-only model of the electron spin qubit to illustrate spin
precession in the non-unidirectional field.
 The Bloch vector changes direction rapidly, which in turn causes the
 Stern-Gerlach force to fluctuate during the operation of the
 gate.
 \label{BlochVecOsc}
 }
\end{figure}

Another source of problems for this field configuration is that
the $|\!\uparrow_y\rangle,\,|\!\downarrow_y\rangle$ states are not
eigenstates of the field as they were with the linearly
inhomogeneous field. Figure~\ref{BlochVecOsc} shows the evolution
of the Bloch vector in a spin-only model of the electron spin
qubit. It demonstrates how the Bloch vector begins precessing
about the $x$ component of the magnetic field, causing the
Stern-Gerlach force to act in different directions. Although
careful timing of the gate and possible compensation mechanisms
could be put in place, this gate is essentially not robust. We
draw the conclusion that in parameter regimes relevant to
single-electron transport by SAW in GaAs, a field of this type
will not be suitable for a Stern-Gerlach measurement gate, and
that field unidirectionality is generally an important requirement
for quantum gates driven by static magnetic fields.

\section{Decoherence effects \label{Section:Neglect}}
The major effects of decoherence on this SAW quantum computing
system have already been considered qualitatively in the original
proposal and subsequent
papers,\cite{Barnes:PRB:2000,Robinson:PRB:2001} and more
specifically by a number of authors. These effects include: The
interaction of the electron qubit with other electrons in the
2DEG, surface gates and donor
impurities;\cite{Robinson:PRB:2001,Barrett:PRB:2003} the coupling
of qubits to phonons,\cite{Golovach:2003,Semenov:PRL:2004} nuclear
spins\cite{Burkard:PRB:99} and radio-frequency
photons\cite{Yamamoto:Book:1999}.

The models we have presented here give rise to behavior which was
previously neglected in the original proposal for the quantum
computation scheme.\cite{Barnes:PRB:2000}  They allow us to
predict the contribution to decoherence from excitation into
higher orbital states and give their probabilities of occupation.
Any potential decoherence due to tunneling between neighboring
dots is negligible, and if needed the spatial separation of qubits
can be increased by either introducing higher harmonics of the SAW
fundamental frequency or by active gating at the entrance to each
Q1DC. A major source of error is the current fluctuations in the
quantized single-electron SAW current. In recent
experiments,\cite{Robinson:PRB:2002,Robinson:Proc:2003} error in
the current quantization (including shot and Johnson noise) of
less than $0.1\%$ were observed.

The relevant decoherence timescales that underpin our proposals is
the spin lifetime ($T_1$ and $T_2$) of single electrons in quantum
dots. For n-type bulk semiconductors, $T_1$ spin lifetimes in GaAs
of $100\units{ns}$ have been reported, which gives a very crude
ball-park figure.\cite{Kikkawa:98} An estimate that is more
relevant to the SAW electron system is the spin relaxation
lifetime of a few nanoseconds, which was obtained from
spin-resolved microphotoluminescence spectroscopy measurements on
photoexcited electrons.\cite{Sogawa:PRL:2001} However, this is the
worst-case scenario because the method of electron capture we
propose\cite{Barnes:PRB:2000} produces a conduction-band hole that
is more extended, short-lived and mixed than the photoexcited
hole. Indeed, recent experiments on static GaAs quantum dots
report a lower bound on $T_1$ as long as $50\units{\mu
s}$,\cite{Hanson:PRL:2003} which makes an estimate of
$100\units{ns}$ for the $T_1$ lifetime of SAW electrons quite
reasonable in the light of current technology.

\section{Summary}\label{Section:Summary}
We have analyzed in detail a proposal\cite{Barnes:PRB:2000} for
implementing quantum computation on electron spin qubits trapped
in SAW-Q1DC electrostatically-defined dots, using only static
magnetic gates to perform single-qubit initializations, rotations
and readouts. Applying the full Pauli Hamiltonian, we described
the quantum dynamics of both the spin and orbital states of the
qubit for various parameter regimes relevant in SAW
single-electron transport experiments. In the analysis of
single-qubit unitary operations with localized uniform magnetic
fields, we showed that the effect of the Lorentz force puts upper
bounds on field strengths and gate times. Moreover, simulations
showed that field directions normal to the 2DEG excite rotational
states in the dot. Probabilities of excitation into higher-energy
orbitals were given. In terms of feasibility, the models indicate
that a field strength of about $80\units{mT}$ will be sufficient
to conduct a $\pi$ rotation in about $1\units{ns}$, without
compromising the confinement properties of the trapped qubit.
Since $T_1$ spin lifetimes of microseconds have been
reported,\cite{Hanson:PRL:2003} it is feasible for hundreds of
single-qubits gates to operate before all coherence is lost in the
computation.

We also studied a device for single-qubit measurement and
initialization based on the Stern-Gerlach effect. The problematic
Lorentz force can be partly suppressed by virtue of the geometry
of the 2DEG system, and for a unidirectional, linearly
inhomogeneous field, the correlation between the spin states and
spatial location of the qubit leads to a good quantum measurement
of its spin. For a 2D dipole field, the vector potential has a
deleterious effect. Namely, it contributes an effective confining
potential via the $A^2$ term in the Pauli Hamiltonian which
suppresses the transverse motion in the $y$ direction.
Furthermore, the component of magnetic field in the $x$ direction
causes the Bloch vector to rotate in undesired directions. The
latter problem could be overcome by using unitary gates prior to
the Stern-Gerlach gate to correct any spurious rotations, but it
is unlikely that such a finely-tuned arrangement can be made
robust. Magnetic fields that are good approximations to a
unidirectional and linearly inhomogeneous field can provide a
source of polarized electron qubit states in channels $C_\uparrow$
and $C_\downarrow$ with high yield. The very same gate can be used
to measure the ratio $|\alpha|^2/|\beta|^2$ for an input qubit
state
$|\psi\rangle=\alpha|\!\uparrow_y\rangle+\beta|\!\downarrow_y\rangle$.
The advantages of this readout method are that the averaging time
of the measured current can be made long enough to render most
noise sources (e.g. shot noise, input-referred noise of the
current pre-amp) unimportant, and that missing electrons from SAW
minima do not contribute any error since we are measuring the
ratio of currents out of the two channels. Field gradients in the
range $0.1-100\units{T\mu m^{-1}}$ lead to gate times that are
easily within the range of $T_1$ spin lifetimes in GaAs. These
parameter regimes will need to be probed by experiment if we are
to make progress towards feasible quantum computation with
nanomagnets in the SAW quantum computer.

\section*{Acknowledgements}
The authors would like to thank Sean Barrett, Masaya Kataoka,
Alexander Moroz, Andy Robinson and Tom Stace for their ideas and
helpful discussions. The authors gratefully acknowledge the
support of the EPSRC. This work was partly funded by the
Cambridge-MIT institute.

\appendix
\section{Derivation of the 2D dipole field vector
potential}\label{DeriveDipoleField} We adopt cylindrical
coordinates $(r,\phi,y)$ about the $y$ axis, with
$r=\sqrt{x^2+z^2}$. We consider the field due to an infinite
string of dipoles passing parallel to the $z$-axis through $y = d$
as shown in Fig.~\ref{VectorField}. The resulting magnetic field
has planar symmetry with respect to the $xy$ (2DEG) plane. The
Q1DC runs along the $x$ axis. A vector potential for this field
can be constructed as follows. A magnetic dipole at $y=z=0$
pointing in the $+y$ direction has a vector
potential\cite{Jackson:EM}
\begin{equation}
 A'_{\phi}(r,y) = \frac{\mu_0 m_B dz}{4\pi} \frac{r}{(r^2+y^2)^{3/2}},
\end{equation}
where $A'_r= A'_y=0$ and $m_B$ is the dipole moment per unit
length in the $z$ direction. We now displace this potential by $d$
in the $y$ direction and integrate over all $z$, obtaining
\begin{equation}
A_z =  \frac{\beta x}{x^2 +(y-d)^2},
\end{equation}
with $A_x=A_y=0$ and strength parameter $\beta=\mu_0 m'_B/2\pi$.
The magnetic field can be derived by applying the curl in
cylindrical coordinates: $B_r = -r^{-1}\partial_y (r A_\phi)$ and
$B_y = r^{-1}\partial_r (rA_\phi)$. Figure~\ref{VectorField} shows
the vector field plot of the resulting magnetic field, which falls
off as $r^{-2}$ for large $r$.

%\bibliographystyle{apsrev}
%\bibliography{../../../LaTeX/master}
%\end{document}

\end{document}